\documentclass[useAMS,usenatbib,12pt,psfig]{mn2e} %,referee
\usepackage{color}
\usepackage{graphicx}
\usepackage{rotating}
\usepackage{txfonts}
\usepackage{pdflscape,lscape}

\newcommand{\hi}{H\,{\sc i}}

\newcommand{\km}{km\,s$^{-1}$}
\newcommand{\degree}{$^{\circ}$}
\newcommand{\halpha}{H${\alpha}$}

\newcommand{\msolar}{M$_{\odot}$}

\newcommand{\mhi}{$M_{\mathrm {HI}}$}

\newcommand{\msolaryr}{M$_{\odot}$\,yr$^{-1}$}
\newcommand{\sett}{SE H\,{\sc i} tidal tail}
\newcommand{\nwtt}{NW H\,{\sc i} tidal tail}
\newcommand{\nwtb}{NW  H\,{\sc i} tidal bridge}
\newcommand{\nwdr}{diffuse NW H\,{\sc i} region}

\title{\hi, star formation and tidal dwarf candidate in the Arp\,305 system}

\author [Sengupta {\it{et al.}}]{Chandreyee Sengupta,$^{1,2}$\thanks{e-mail:sengupta.chandreyee@gmail.com(CS), tom.scott@astro.up.pt(TS), dwaraka@rri.res.in(KSD), djs@ncra.tifr.res.in(DJS), bws@kasi.re.kr(BWS)} T. C. Scott$^{3,4}$, S. Paudel$^{2}$, K. S. Dwarakanath$^{5}$, D. J. Saikia$^{6,7}$ 
\newauthor and B. W. Sohn$^{2}$  \\ \\
$^{1}$ Department of Astronomy, Yonsei University, 50 Yonsei-ro, Seodaemun-gu, Seoul, Republic of Korea\\
$^{2}$ Korea Astronomy and Space Science Institute, 776, Daedeokdae ro, Yuseong gu, Daejeon, 305-348, Republic of Korea\\ 
$^{3}$ Institute of Astrophysics and Space Sciences (IA), Rua das Estrelas, 4150-762 Porto, Portugal\\
$^{4}$ Centre for Astrophysics Research, University of Hertfordshire, College Lane, Hatfield, AL10 9AB, UK \\
$^{5}$ Raman Research Institute, Bangalore 560 080, India \\
$^{6}$ National Centre for Radio Astrophysics, Tata Institute of Fundamental Research, Pune 411 007, India \\
$^{7}$ Cotton College State University, Panbazar, Guwahati 781 001, India \\}

\begin{document}

\date{Received  ; accepted  }
\date{}
\pagerange{\pageref{firstpage}--\pageref{lastpage}} \pubyear{}

\maketitle

\label{firstpage}

\begin{abstract}
 We present results from our  Giant  Metrewave Radio Telescope  (GMRT) \hi\ observations of the Arp\,305 system. The system consists of two interacting spiral galaxies NGC\,4016 and NGC\,4017, \textcolor{black}{a} large amount of resultant tidal debris and a prominent tidal dwarf galaxy (TDG) candidate projected within the tidal bridge between the two principal galaxies. Our  higher resolution  GMRT \hi\ mapping,   compared to previous observations,  allowed detailed study of smaller scale  features.   Our \hi\ analysis   supports the conclusion in \cite{hancock09}  that the  most recent encounter between the pair occurred $\sim$ 4 $\times$ 10$^8$ yrs ago\textcolor{black}{. The GMRT observations also show} \hi\ features near NGC\,4017 which may be remnants of  an earlier encounter between the two galaxies. The \hi\  properties of the Bridge TDG candidate \textcolor{black}{include:} \textcolor{black}{\mhi\ }  $\sim$  6.6 $\times$ 10$^8$\msolar\ and V$_{HI}$ = 3500$\pm$ 7\km\textcolor{black}{, which is in good agreement with the velocities of the parent galaxies}. Additionally \textcolor{black}{the TDG's \hi\  linewidth of 30 \km\ and   modest velocity gradient  together with its} SFR of 0.2 \msolaryr\  add to the evidence favouring  the bridge  candidate being  a genuine TDG.  \textcolor{black}{The Bridge TDG's \textit{Spitzer} 3.6 $\mu$m  and 4.5 $\mu$m  counterparts with a [3.6]--[4.5] colour $\sim$ -0.2 mag  suggests stellar debris may have seeded its  formation}. Future spectroscopic observations could confirm this formation scenario and provide  the metallicity which is a key criteria  for the validation for TDG candidates.

\end{abstract}

\begin{keywords}
galaxies: spiral - galaxies: interactions - galaxies: kinematics and dynamics - 
galaxies: individual: Arp\,305 - radio lines: galaxies - radio continuum: galaxies \textcolor{black}{galaxies: tidal dwarfs}
\end{keywords}

 \section{Introduction}
\label{intro}
 \textcolor{black}{Interactions between gas rich galaxy pairs, can result in massive \hi\ stripping from the parent \textcolor{black}{galaxies' \hi\ disks}  \textcolor{black}{\citep[e.g.,][]{duc97,duc00,smith07,smith10, seng15}}. Much of this stripped \hi\ may fall back into the \textcolor{black}{gravitational} potential of either of the pair or, at later stages, the new merged galaxy or be incorporated into the intra--group medium (IGM). } Under the right conditions new star formation \textcolor{black}{ may arise within the evolving \hi\ debris \textcolor{black}{\citep{hibbard05,neff05,smith10, demello12,torres12}}.  \textcolor{black}{If the cold gas (\hi\ and molecular gas) densities are sufficient and environmental conditions are favourable, the evolution of the cold gas and stellar debris may include the formation of self--gravitating bodies with masses typical of dwarf galaxies, known as Tidal Dwarf Galaxies (TDG)}  \citep{duc99,duc00,smith07,smith10}. }   \textcolor{black}{\halpha\ and \textcolor{black}{UV emission} trace star formation on timescales of 10$^7$\,yr and 10$^8$\,yr respectively \citep{bosel09}.  \textcolor{black}{So\textcolor{black}{,} if a TDG has an \hi\  counterpart and  it has stellar populations formed later than the time when its parent galaxies began interacting, this can provide evidence for in-situ star formation.}}  \textcolor{black}{Cases where a TDG forms from pure gas collapse, as opposed to gas collapse \textcolor{black}{promoted by the gravitational potential of }  stellar debris from the parents, can be considered as a separate class of TDG \citep{duc04}.} Interacting pairs  thus provide a unique  \textcolor{black}laboratory to study the impact of  interactions on the gaseous and the stellar components of the parent galaxies as well as the conditions  under which neutral gas debris collapses to form star clusters and TDGs.

%Much of the effort to validate  the existence of TDG candidates  is directed at to differentiating their properties those of  the normal  dwarf galaxies  commonly associated with large galaxies. We are only in the initial stages formulating tests to differentiate TDGs from normal dwarfs.

\begin{figure*}
\begin{center}
\includegraphics[ angle=0,scale=.50] {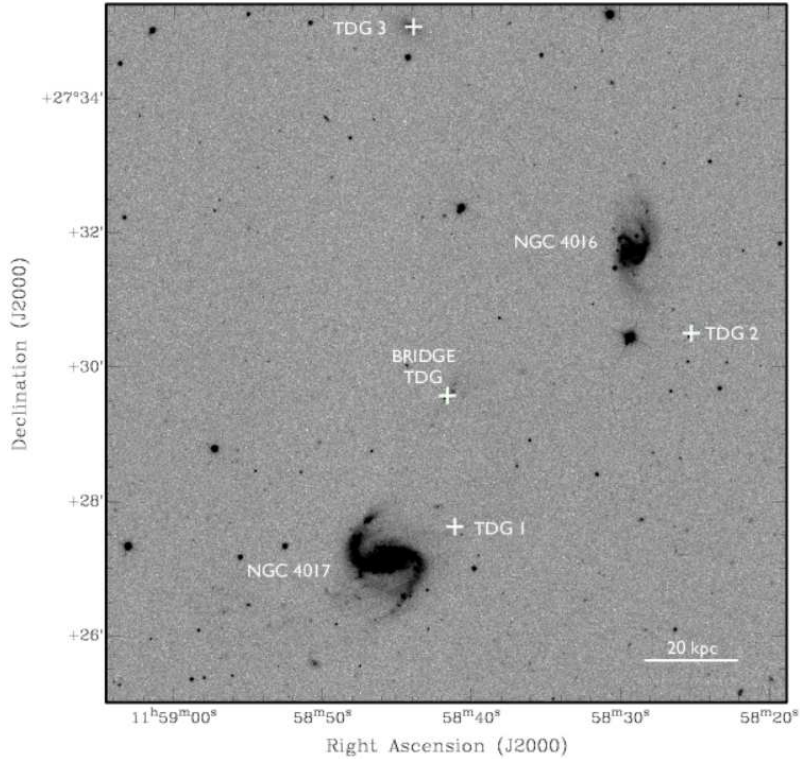}
\vspace{1cm}
\caption{\textbf{Arp\,305:} SDSS g -- band  image, with the positions of \textcolor{black}{the} two principal galaxies, NGC\,4016 and  NGC\,4017, indicated as well as the positions of the four TDG candidates from \citep{hancock09}.  }
\label{fig1}
\end{center}
\end{figure*}

%A multi--wavelength study of a sample of Arp interacting galaxies (\textcolor{black}{The} ‘Spirals, Bridges, and Tails’ (SB\&T) sample) has been carried out \textcolor{black}{\citep{smith07,smith10} using UV, NIR}  and optical  observations. Their studies provide evidence of in--situ star formation and TDG candidates in several interacting systems, \textcolor{black}{ including} Arp\,305.  
Arp\,305 is an interacting pair of galaxies (NGC\,4016 and  NGC\,4017) with a M$_*$ ratio of $\sim$ \textcolor{black}{1:3} and  heliocentric optical radial velocities\footnote{ From Hyperleda \citep{makarov14}.} of 3441$\pm1$ \km\ and 3449$\pm2$ \km\ respectively. \textcolor{black}{Further basic properties of the pair are set out in Table \ref{table_1}.}    NGC\,4016 and  NGC\,4017  are part of  a small group of 5 galaxies (USGC U435)  with its centre projected at \textcolor{black}{11:58:17.8	+27:47:03} with a radial velocity of	3456 \km\ \citep{ramella02}. \textcolor{black}{ The group velocity dispersion  is 106 \km.}  The closest  member of the group \textcolor{black}{ is projected } $\sim$ 22 arcmin (341 kpc) north of Arp\,305. At optical and UV wavelengths, the  \textcolor{black}{Arp\,305} pair displays clear  signatures { of \textcolor{black}{a tidal interaction}. These  signatures  include a figure of eight shaped inner disk in \textcolor{black}{NGC\,4016,} enhanced spiral arms in NGC\,4017 \textcolor{black}{and a tidal bridge remnant projected between the pair \textcolor{black}{(Figure \ref{fig1})} as well as four TDG candidates \citep{hancock09}.}

% with total masses of 1.7$\times$10$^{10}$ \msolar\ and 16.2$\times$10$^{10}$ \msolar\ respectively \citep{hancock09}.     projected 5.9 arcmin (93 kpc) from each other 

% an enhanced spiral arm eminating from the northen end of the NGC\,4017 bar then chanong direction by  $\sim$ 90\degree at the optical disk edge and becoming the  SE tidal tail,    Based on  modeling \textcolor{black}{of the stellar component of the pair }  \cite{hancock09}  concluded that the pair had a  prograde grazing encounter  $\sim$ 300 Myr ago. The  9 $\pm$2 \km\ difference in the optical radial velocities and 356 arcsec (86  kpc)  projected separation between the pair suggest the interaction axis lies principally in the plane of the sky. Assuming this correct the projected separation implies an interaction velocity of $\sim$ 320 \km, in good agreement with the 300 \km\  from the \cite{hancock09} modeling. 

A previous  \hi\ mapping of \textcolor{black}{Arp\,305  by} \cite{vanmoor83} \textcolor{black}{with the Westerbork Synthesis Radio Telescope (WSRT)} detected \hi\ in both members of the pair and } indicated they are at an early stage of a wet merger. \cite{hancock09}  used UV (\textit{GALEX}) observations to identify 45 young star forming clumps, including clumps within \textcolor{black}{ the}  four TDG candidates  (see Figure \ref {fig1})\textcolor{black}{. The} ``Bridge TDG" candidate corresponds to the UV \textcolor{black}{clumps} 12, 13,15 and 16 in Figure 2 of \cite{hancock09}, with TDG1, TDG2 and TDG3 correspond to UV clumps 11, 1, and 19 respectively in the same figure.  

In this paper, we present results from our  Giant  Metrewave Radio Telescope  (GMRT) \hi\ observations of the Arp\,305 system. These \textcolor{black}{observations have a  higher spatial and velocity resolution than the previous WSRT \hi\ observations } \citep{vanmoor83}  \textcolor{black}{allowing the} detailed study of small scale \hi\ morphology and kinematic features within the Arp\,305 system\textcolor{black}{, and in particular the star forming regions}.  This paper also utilises  \textcolor{black}{the Sloan Digital Sky Survey SDSS}, \textit{Spitzer}  and  \textcolor{black}{Galaxy Evolution Explorer (\textit{GALEX})} publicly available \textcolor{black}{archive} data and images.  Section \ref{obs} sets out details of our observations, \textcolor{black}{with observational  results} given in section \ref{results}. We discuss the results in section \ref{dis}. A summary and concluding remarks are set out in section \ref{summary}. Using the average heliocentric velocity of the two principal galaxies from Hyperleda and assuming H$_{0}$ to be \textcolor{black}{68 km s$^{-1}$ Mpc$^{-1}$\citep{ade14}, we adopt a distance of 50 Mpc} for NGC\,4016 and NGC\,4017 and the TDGs. At this distance the spatial scale   is $\sim$ 14.5 kpc/arcmin. \textcolor{black}{These values are comparable to those used by  \textcolor{black}{\cite{hancock09} and} \cite{smith10}. } J2000 coordinates are used throughout the paper, including in the figures.

%As noted above  \textcolor{black}{ \hi\ has  previously been mapped in the Arp\,305 system with the  WSRT\footnote{Westerbork Synthesis Radio Telescope } \citep{vanmoor83} and  established the presence of  large amounts of tidal  \hi\ debris in the system. The  shortest  baseline spacing for the WSRT was shorter  than for the GMRT observations, allowing  the WSRT to recover a larger fraction of the \hi\  flux. However the  spatial resolution of the \cite{vanmoor83} observations ($\sim$  55$^{\prime\prime}$) was poorer than both the GMRT high ($\sim$ 30  $^{\prime\prime}$) and low ($\sim$ 20$^{\prime\prime}$) resolution  data. 

%Moreover the WSRT  velocity resolution  (16.9 \km) was less than half of the 7 \km\ available from the GMRT observations. 

% The improvement in spatial and velocity resolution from the GMRT data 

%  In particular this using the WSRT. However the significantly higher \textcolor{black}{spatial and velocity  resolution [and sensitivity??]} of our GMRT  observations allows us to  study the  relationship between  recent star forming clumps and their   hi\ counterparts.} 

%The ellipse at the bottom left indicates the size and orientation of the GMRT synthesised beam. The thick white contours are the  8 $\sigma$ contours from the FUV (\textit{GALEX}) image. The ellipse at the bottom right shows the size and orientation of the low resolution beam.[THE MOMENT 0 FIGS 1 and 2 HAVE A DIFFERENT SENSITIVITY CUT OFF TO THIS MOMENT 1 MAP -- 2 or 3 $\sigma$ ?]

\begin{table}
\centering
\begin{minipage}{190mm}
\caption{\textcolor{black}{Properties of the Arp\,305 pair}}
\label{table_1}
\begin{tabular}{llrr}
\hline
%Frequency & Observation  & Phase      & Phase cal    &  $\tau$    & Bandwidth &rms (per channel  & beam size   \\ 
Property\footnote{All data  are from NED, except V$_{radial(optical)}$ and Inclination\\
which are from \textcolor{black}{Hyperleda.}}&Units&NGC\,4016&NGC\,4017 \\ 
\hline
V$_{radial(optical)}$&[\km]& \textcolor{black}{3441$\pm$1} &\textcolor{black}{3449$\pm$2}\\
RA&[h:m:s]&\textcolor{black}{11:58:29.02}&\textcolor{black}{11:58:45.67}\\
DEC&[d:m:s]&\textcolor{black}{+27:31:43.62}&\textcolor{black}{+27:27:08.79}\\
Distance\footnote{\textcolor{black}{See section 1}.}&[Mpc]&  \textcolor{black}{50} &  \textcolor{black}{50} \\
%&& (\textcolor{black}{accepted})\\
%Spatial scale&[kpc/arcmin]&20.94 & 21.04 &NED \\
D$_{25}$ major /minor  &[arcmin]&  \textcolor{black}{1.5 x 0.8} & \textcolor{black}{1.8 x 1.4}  \\
D$_{25}$ major /minor& [kpc]& \textcolor{black}{21.8 x 11.6} & \textcolor{black}{26.1 x 20.3}\\
Inclination& [\degree] & \textcolor{black}{59.8}  & \textcolor{black}{48.2}   \\
Morphology&&\textcolor{black}{SBdm}&\textcolor{black}{SABbc}  \\
%\hi\ \aflux&&1.08 & \\
B$_T$& [B band mag]&14.54$\pm$ 0.13   &14.34$\pm$0.13   \\
%log($L_B$)&  [\lsolar] &14.66&14.78  \\
Stellar mass $M_*$  &[10$^{10}$ \msolar] &  \textcolor{black}{0.6} &  \textcolor{black}{2.7.} \\
%log(L$_{FIR}$) & [\lsolar]&9.53& \\
%%\hline
\end{tabular}
\end{minipage}
\end{table}

\section{Observations}
\label{obs}
\hi\  observations of Arp\,305 were carried out with the GMRT on \textcolor{black}{July 12th, 2014}.  A baseband bandwidth of 16 MHz was used  for the \hi\ 21-cm  line observations. The resultant velocity resolution was $\sim$\textcolor{black}{7} km~s$^{-1}$. \textcolor{black}{Further details of the observations  are given in Table \ref{table2}.}
 
The  Astronomical Image Processing System ({\tt AIPS}) software package was used to reduce the data. Data  from malfunctioning antennas, low gain antennas and/or \textcolor{black}{antennas} suffering from radio frequency interference (RFI) were flagged.   The flux density calibration scale used was \cite{baars77}, with flux density uncertainties $\sim$5\%.  After calibration, the uv domain continuum subtraction was \textcolor{black}{carried out}  using the {\tt AIPS} task \textsc{uvlin}. The task \textsc{imagr} was then \textcolor{black}{applied to the visibilities  to `clean' and transform them into \hi\ \textcolor{black}{image} cubes.}   The \textcolor{black}{ integrated \hi,  \hi\   velocity field and velocity dispersion }  maps were made applying the {\tt AIPS} task \textsc{momnt} on the \hi\ cubes. To analyse the detailed \hi\ morphology and kinematics,  images \textcolor{black}{with} different resolutions were produced by applying different \textcolor{black}{ `tapers' to the data with varying} uv limits. \textcolor{black}{Details of the final low and high  resolution map properties  are given in Table \ref{table2}.}

\begin{table}
\centering
\begin{minipage}{110mm}
\caption{GMRT observation details}
\label{table2}
\begin{tabular}{ll}
\hline
Frequency & 1420.4057 MHz \\
Observation Date &\textcolor{black}{12th July, 2014 }\\
Primary calibrator&3C147\\ 
Phase calibrator  & \textcolor{black}{1120+143 (2.42 mJy) } \\
(flux density)  &   \\
Integration time  & \textcolor{black}{10.0 hrs } \\
primary beam & 24\arcmin ~at 1420.4057 MHz \\
Low resolution beam   &\textcolor{black}{ 31.8$^{\prime\prime}$ $\times$ 29.5$^{\prime\prime}$  (PA = -0.8$^{\circ}$)} \\
%Medium resolution beam & \textcolor{black}{23.4}$^{\prime\prime}$ $\times$ \textcolor{black}{ 20.5}$^{\prime\prime}$ (PA = \textcolor{black}{21.5}$^{\circ}$) \\
High resolution  beam   & \textcolor{black}{14.3}$^{\prime\prime}$ $\times$ \textcolor{black}{11.8}$^{\prime\prime}$ (PA = \textcolor{black}{27.8}$^{\circ}$) \\
rms for low resolution map  & \textcolor{black}{1.2 mJy beam$^{-1}$ } \\
%rms for medium resolution map & 0.7 mJy beam$^{-1}$  \\
rms for high resolution map & 0.6 mJy beam$^{-1}$  \\
RA (pointing centre)&\textcolor{black}{ 11$^{\rm h}$ 58$^{\rm m}$ 37.4$^{\rm s}$  }\\
DEC (pointing centre)& \textcolor{black}{27$^\circ$ 29$^\prime$ 26$^{\prime\prime}$.9}\\
\hline
\end{tabular}
\end{minipage}
\end{table}

\section{Observational Results} 
\label{results}

\subsection{\hi\ \textcolor{black}{morphologies}  and mass estimates} 
\label{res_morph_mas} 
\begin{figure*}
\begin{center}
\includegraphics[angle=0.0,scale=.45] {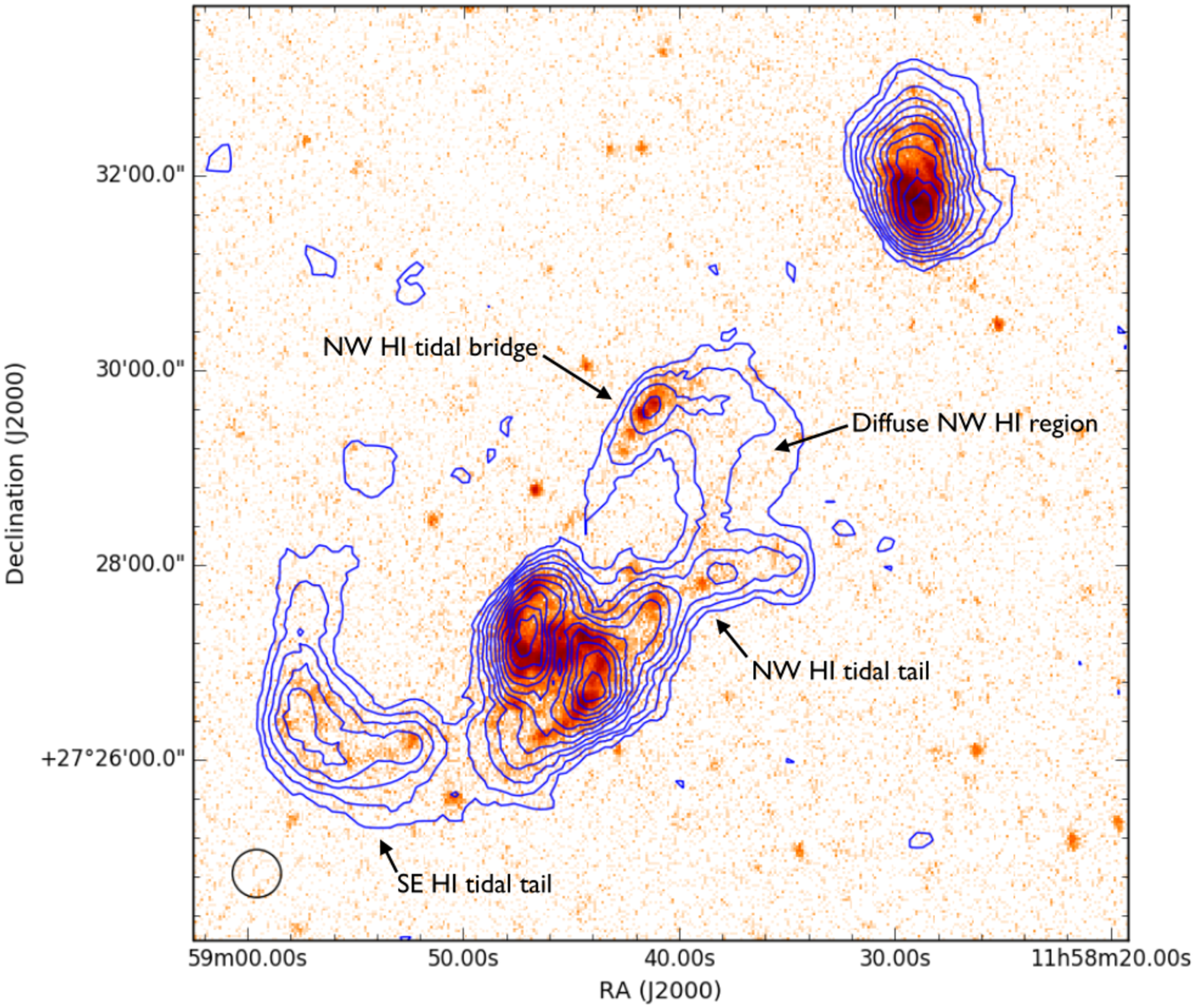}
\vspace{1cm}
\caption{\textbf{Arp\,305:  Integrated \hi\ contours from the GMRT low resolution map } overlaid on a FUV (\textit{GALEX}) image.  The  \hi\ column density contour levels are  N$_{HI}$ = 10$^{20}$ atoms cm $^{-2}$ (0.5, 1.7, 2.9, 4.1, 5.3, 6.4, 8.8, 11.1, 12.9). Major \hi\ tidal features referred in the text  are marked. The ellipse at the bottom left shows the size of  low resolution   \textcolor{black}{(31.8$^{\prime\prime}$ $\times$ 29.5$^{\prime\prime}$)} synthesised beam.} 
\label{fig2}
\end{center}
\end{figure*}

Figure \ref{fig2} shows the contours from the  GMRT low resolution (31.8$^{\prime\prime}$ $\times$ 29.5$^{\prime\prime}$)  integrated \hi\  map for the Arp\,305 field  overlaid on a  FUV (\textit{GALEX}) image. The bulk  of  the \hi\ is detected  in  NGC\,4017,  with significant amounts of its \hi\ detected at lower column densities (0.5 -- \textcolor{black}{4.3 $\times$ 10$^{20}$} cm$^{-2}$) in \textcolor{black}{an}  extended south eastern (SE)  tidal tail  and a broad \textcolor{black}{area north (N)} and \textcolor{black}{Northwest (NW) of the } optical  disk.  \textcolor{black}{We refer to \hi\ column densities as `lower' or `higher' with respect to the  \hi\ column density threshold of $\sim$ 4 $\times$ 10$^{20}$ atoms cm $^{-2}$ for star formation, as estimated by \cite{maybhate07}}.  

To a first order the  morphologies and the peak \hi\ column densities of both members of the Arp\,305 pair in the GMRT low resolution (synthesised beam $\sim$ 31.8$^{\prime\prime}$ \textcolor{black}{$\times$} 29.5$^{\prime\prime}$) and  lower resolution  WSRT \citep{vanmoor83} (synthesised beam $\sim$ 45$^{\prime\prime}$ \textcolor{black}{$\times$} 60$^{\prime\prime}$)  \hi\ maps  are  similar. \textcolor{black}{Although it is evident from a comparison of the GMRT and WSRT  \hi\ morphologies  that the GMRT has suffered some flux loss due to a lack of short spacing baselines.  \textcolor{black}{ For NGC\,4016, Figure \ref{fig2} shows its  \hi\ is truncated  in  the south to approximately the optical disk radius with the \hi\ column densities rising  rapidly toward the column density maximum. In the north the \hi\ disk extends beyond the optical disk.     No \hi\ counterparts to  TDG\,2 and TDG\,3 were detected by the GMRT.  In the case of  NGC\,4017, Figure \ref{fig2}  shows  massive extended \hi\ tidal tails  SE and NW of the galaxy's FUV  disk.  The SE tidal tail (henceforth the ``\sett")  contains  high column density  \hi, within which are projected  several star forming (SF) clumps detected in UV}.  North (N) and NW  of the NGC\,4017 optical disk the \hi\ morphology is more complex.  Figure \ref{fig2}  shows it contains two elongated  \hi\ structures  of relatively high column density with FUV counterparts. The first structure is an \hi\ extension,  with a clumpy FUV counterpart,  including  TDG\,1, running  from the western edge of the optical/FUV  disk  to the   NW \textcolor{black}{(henceforth the ``\nwtt")}.  The second prominent \hi\ structure  is the \hi\ counterpart to the UV and optical  tidal bridge remnant between NGC\,4017 and NGC\,4016, referred to from here on as the  `` \nwtb". The  Bridge TDG candidate is projected within the \nwtb. Both of these large scale \hi\ structures, also visible in the WSRT map,  are embedded within a much more extensive  lower \hi\ column density region,  which in the WSRT map  extends $\sim$ 1.5 arcmin (22 kpc) further north  than in GMRT map,  almost connecting to the NGC\,4016 \hi\ disk. This region is referred to as the ``\nwdr".  Properties of the two principal galaxies and the TDG candidates, including their GMRT \hi\ masses \textcolor{black}{as well as their velocities} and W${_{20}}$ line widths, are set out in Table  \ref{table4}.} 

 \textcolor{black}{Comparing} the \hi\  masses for NGC\,4016 and   NGC\,4017 derived from \textcolor{black}{the GMRT, to their}  literature \hi\ flux densities \textcolor{black}{is} complicated by the following factors:  \textcolor{black}{(i) there are conflicting} \hi\ flux densities reported in the literature for the galaxies from several  single dish and a single WSRT interferometric observation \citep{vanmoor83} \textcolor{black}{(ii)} the close proximity of NGC\,4016 and   NGC\,4017 and large single dish beam sizes mean \hi\ emission is likely to be partially confused within the single dish beams. Only the GMRT  and   WSRT \hi\ maps  resolve  NGC\,4016 and NGC\,4017 into discrete \hi\ entities.   \textcolor{black}{(iii)} It is difficult to accept that the  WSRT flux  calibration was accurate because  the 1.4 GHz WSRT  radio continuum flux density for NGC\,4016 was $\sim$ 1.5 times the NRAO VLA Sky Survey (NVSS) value but for  NGC\,4017 the WSRT radio continuum flux density is similar to NVSS value.  \textcolor{black}{(iv)} \textcolor{black}{as} noted above, comparison of the WSRT and GMRT \hi\ maps indicates the GMRT observation did suffer some flux loss. Below we compare the GMRT and literature   \hi\ flux densities values  for NGC\,4016 and   NGC\,4017. 

\textcolor{black}{The} GMRT integrated \hi\ flux density (\textit{S}) for NGC\,4016 was \textit{S} = 5.0 Jy \km\ compared to  \textit{S} = 7.5 Jy \km\ from the \cite{vanmoor83} \textcolor{black}{WSRT } \hi\ mapping.   Single dish flux density measurements for NGC\,4016 have been reported: \textit{S} = 10.05 Jy \km\ using the 305 m Arecibo telescope  \citep{haynes11}; and \textit{S} = 8.1 Jy \km\ from the Nan\c{c}ay radiotelescope \citep{theureau07} and 6.2 Jy \km\ from Arecibo telescope \citep{1989gcho.book.....H}.  The  Nan\c{c}ay beam  is quite large (FWHP   3.6 arcmin  $\times$ 22 arcmin  at zero declination). Assuming the Nan\c{c}ay observation was centred on NGC\,4016, the flux density derived in  \cite{theureau07} would be contaiminated with emission from  NGC\,4017, with the severity depending  on the size and orientation of the beam.   Arecibo's  beam has a FWHP $\sim$3.5 arcmin. \cite{haynes11} note the  \hi\ emission detected with Arecibo at the position of NGC\,4016 is ``probably" blended with emission from NGC\,4017 and \textcolor{black}{their} reported flux density for NGC\,4016 is after  ``attempted" deblending. \textcolor{black}{Since the \hi\ flux values reported in the literature were so diverse, we compared our GMRT 20 cm radio continuum flux value of NGC\,4016, (5.3 mJy$\pm$ 10\%) with that  \textcolor{black}{from the NVSS} (5.5 mJy$\pm$ 10\%) as a check for our GMRT calibration. \textcolor{black}{ These matching continuum values indicate that our calibration is accurate and the literature \hi\  flux densities   which differ widely from the GMRT value for NGC\,4016 cannot be accepted as reliable} measurements.}% Nonetheless, both for NGC\,4016 as well as NGC\,4017, we \textcolor{black}{quote} here the \hi\ \textcolor{red}{flux density} measurements found in the literature and discuss possible reasons for  the large discrepancies between  them. }

 \textcolor{black}{For NGC\,4017 the GMRT integrated interfrometric flux density is \textit{S} =  8.5 Jy \km\ compared to \textit{S} =  25.8 Jy \km\  from the WSRT \citep[][]{vanmoor83}. \textcolor{black}{In part the difference is attributable to the GMRT \hi\ flux loss referred to above and \textcolor{black}{the} similarity of  the peak \hi\ column densities in  the WSRT and GMRT maps are consistent with this. However, the NGC\,4017 \cite{vanmoor83} \hi\ flux density (\textit{S} = 25.8 Jy \km)} is in good agreement with that derived by  \cite{haynes11} from Arecibo single dish obervations. But, \cite{haynes11} note the possible blending of \hi\ emission from NGC\,4016.  A \textcolor{black}{significantly lower Arecibo based \hi\ flux density for NGC\,4017 of \textit{S} = 16.8  Jy \km\ is reported in } \cite{lewis85}. The NGC\,4017 \hi\ extent   is greater than a single Arecibo $\sim$ 3.5 arcmin  beam, thus requiring integration of the flux densities from multiple pointings. \textcolor{black}{ But neither \cite{haynes11} nor \cite{lewis85} state} which beam areas were used to derive \textcolor{black}{their} reported NGC\,4017 flux densities. }

 \textcolor{black}{ Compared to the GMRT \hi\ flux density, the NGC\,4017 flux densities from \cite{haynes11} and  \cite{vanmoor83}  are about 3 times higher. While some flux loss is expected in GMRT data due to \textcolor{black}{the} lack of short baselines, \textcolor{black}{the flux  loss of $\sim$60\% implied by the \cite{haynes11} and  \cite{vanmoor83} flux densities} is much higher than expected, based on  similar GMRT observations. Furthermore, the \textcolor{black}{``expected" \hi\ mass of a galaxy of NGC\,4017's size and Hubble type, based on a large sample of field galaxies  and using the formula from \cite{hayn84} (\mhi\ = 4.1  $\times$ 10$^{9}$ M$_\odot$),} is  in good agreement the  \hi\ mass derived from the  GMRT (\mhi\ = 4.8  $\times$ 10$^{9}$ M$_\odot$).  The NGC\,4017  \hi\ mass derived from  \cite{haynes11} and  \cite{vanmoor83} flux densities  are also significantly higher than the \textcolor{black}{ \hi\ mass derived from} applying, the Tully--Fisher (TF) relations. Our stellar mass estimate for NGC\,4017 of 3.0 $\times$ 10$^{10}$ M$_\odot$ agrees well with the stellar TF mass relation, based on an \hi\  rotation velocity of $\sim$ 201 \km\  (150 \km\ from the rotating disk adjusted for inclination).  But using the NGC\,4017 \hi\ flux density from \cite{haynes11} and assuming molecular mass = \hi\ mass, gives a baryonic  mass of 6.6 $\times$ 10$^{10}$ M$_\odot$,  significantly greater than the \textcolor{black}{barionic mass for NGC\,4017} from the  TF baryonic mass relation  (3.8 $\times$ 10$^{10}$ M$_\odot$). The TF baryonic mass relation is a tighter relation than the Stellar TF according to \cite{torres}.  Hence the \textcolor{black}{ NGC\,4017 \hi\ mass from the ``expected" and  TF analysis are closer to  those derived from the GMRT and  \cite{lewis85}  \hi\ flux densities than those derivered  from the} \cite{haynes11}  and \cite{vanmoor83} flux densities.   We conclude that the greater extent of the \hi\ detection in the \cite{vanmoor83} map clearly shows the  GMRT data is missing some flux.  But \textcolor{black}{large} uncertainties about reliability of the flux densities in the literature prevent us from quantifying the amount of this loss. Analysis in this paper however concentrates on the relatively compact, high--density  \hi\ star forming zones where flux loss should not affect any of our results significantly. }

\begin{table*} 
\centering
\begin{minipage}{190mm}
\caption{GMRT \hi\ detections}
\label{table4}
\begin{tabular}[h]{@{}lllrrrrl@{}}
\hline
Object& RA& Dec&Velocity\footnote{From \textcolor{black}{GMRT.}}&W$_{20}$ $^a$& \textcolor{black}{\mhi\ }\footnote{M(\hi) for NGC\,4016 and  NGC\,4017 are derived from the GMRT flux densities. \textcolor{black}{But see the caveats for the \hi\ flux densities measured from GMRT as well as those found in the literature in section \ref{res_morph_mas}. } For the Bridge TDG the \hi\ mass was calculated as per section \ref{dis_tdg}. \textcolor{black}{ The other \hi\ detected candidate, TDG\,1, is \textcolor{black}{small UV clump} and embedded in \textcolor{black}{a much larger mass of} \hi\ debris close to the \textcolor{black}{NGC\,4017} disk. Its \textcolor{black}{\hi\ } mass was not estimated due to the high uncertainties in its extent and the corresponding flux density.} } &M$_*$ $^c$&FUV clumps\footnote{\textcolor{black}{From} \textcolor{black}{Table 5 in} \cite{hancock09}, except \textcolor{black}{for} the M$_*$ masses \textcolor{black}{of}  NGC\,4016 and NGC\,4017 which \textcolor{black}{were} calculated using \textcolor{black}{the} method from \cite{bell03}  and parameters from \cite{blanton03} based on the galaxies'   SDSS \textit{r} --band magnitudes and \textit{r-i} colours.  }\\
&[h:m:s]&[d:m:s]  &[\km]&[\km]&[$\times$ 10$^9$ \msolar]&[$\times$ 10$^6$ \msolar]\\
\hline
Arp\,305 pair:& & &&\\
-- NGC\,4016& 11:58:29.0&+27:31:44&\textcolor{black}{3454$\pm$7}&\textcolor{black}{130$\pm$7}&\textcolor{black}{3.0}&\textcolor{black}{10400}&1 to 10\\
-- NGC\,4017&11:58:45.7&+27:27:09&\textcolor{black}{3439$\pm$7}&\textcolor{black}{258$\pm$7}&5.0&\textcolor{black}{33900}&11, 14, 17, 18, 20 to 45  \\
%-- NGC\,4017 SE tidal tail &&\\
%-- NGC\,4017 NW tidal tail &&\\
%-- Tidal tail&\\
Bridge TDG candidate&11:58:42.22 &+27:29:20.44& 3500$\pm$7 &30$\pm$7 &0.66 &1-7&12,13, 15,16\\
%TDG 1 candidate&11:58:41.09&+27:27:38.45& 3524$\pm$7 &30$\pm$7 &--&0.45-3.8&11\\
%TDG 2 candidate&11:58:28.97&+27:31:25.00&--&--&$<$?&19-21&1\\
%TDG 3 candidate&11:58:43.87&+27:35:03.95&--&--&$<$??&6-50&19\\

\hline
\end{tabular}
\end{minipage}
\end{table*}

%\begin{figure*}
%\begin{center}
%\includegraphics[ angle=-90,scale=.65] {LR-MOM0-NWdebris-FUV-NOV2016.eps}
%\vspace{1cm}
%\caption{\textbf{NGC\,4017:} Zoom in on the NW tidal tail and Bridge TDG  region of  the velocity integrated \hi\ map on figure \ref{fig1}.  The first \hi\ contour (blue) is at the 3 $\sigma$ level equivalent to a column density of N$_{HI}$ = x atoms cm $^{-2}$ with the higher level contours at x x atoms cm $^{-2}$. The ellipse at the bottom left indicates the size and orientation of the GMRT synthesised beam.}
%\label{fig3}
%\end{center}
%\end{figure*}

\subsection{ \hi\ kinematics}
\label{results_kin}

Figure \ref{fig4} shows the low resolution \textcolor{black}{($\sim$ 30$^{\prime\prime}$) intensity weighted velocity field of the Arp\,305 system, with  iso--velocity contours separated by 7 \km.} \textcolor{black}{ For \textcolor{black}{NGC\,4016,} \hi\ emission is detected in the channel maps (Figures 8 and 9) within a velocity range  3348 \km\ to 3496  \km. The \textcolor{black}{NGC\,4016}}  iso--velocity contours  (Figure \ref{fig4}) indicate  reasonably regular rotation in the  \hi\  disk, with a north-south (N--S) kinematic axis.  The  closed iso--velocity contours at the \hi\ disk edges indicate the disk is  warped with  the velocity gradient \textcolor{black}{ becoming} progressively shallower towards the north.   NGC\,4016  has \textcolor{black}{a} GMRT  V$_{HI}$ = 3454$\pm$7 \textcolor{black}{\km, }    with   W$_{20}$ $\sim$  130$\pm$7 \km,  close to the   W$_{50}$  $\sim$  133$\pm$15 \km\  from the single dish on--line data from  \cite{haynes11}. 
%n $>$ 2? $\sigma$ [lower sigma in moment 1 than moment 0 map ??]

\begin{figure*}
\begin{center}
\includegraphics[ angle=0,scale=.6] {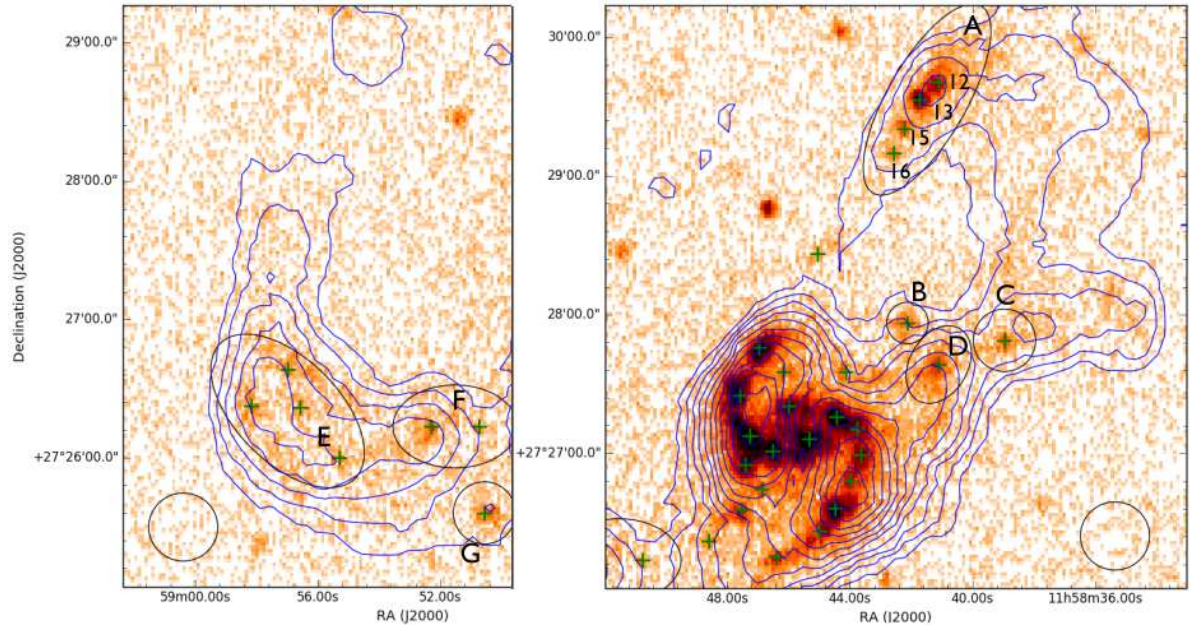}
\vspace{1cm}
\caption{\textbf{NGC\,4017 zoom in \textcolor{black}{on  the low resolution  integrated \hi\ map contours overlaid on an} FUV (\textit{GALEX}) \textcolor{black}{image. }  \textit{Left:}}  \sett. \textcolor{black}{\textbf{\textit{Right:}}} \nwtt\ and \nwtb\  regions of NGC\,4017.  The green crosses mark the positions of the UV clumps from Hancock (2009), with the black \textcolor{black}{ellipses} marking the SF zones A -- G. \textcolor{black}{For SF zone A (The Bridge TDG) the \citep{hancock09} FUV clump number is also indicated.} The \hi\ contour details are as per Figure \ref{fig2}. The ellipses at the bottom of the figures indicate the GMRT \textcolor{black}{low} resolution synthesised beam  \textcolor{black}{( 31.8$^{\prime\prime}$ $\times$ 29.5$^{\prime\prime}$)}. }
\label{fig3}
\end{center}
\end{figure*}

\begin{figure*}
\begin{center}
\includegraphics[ angle=0,scale=.57] {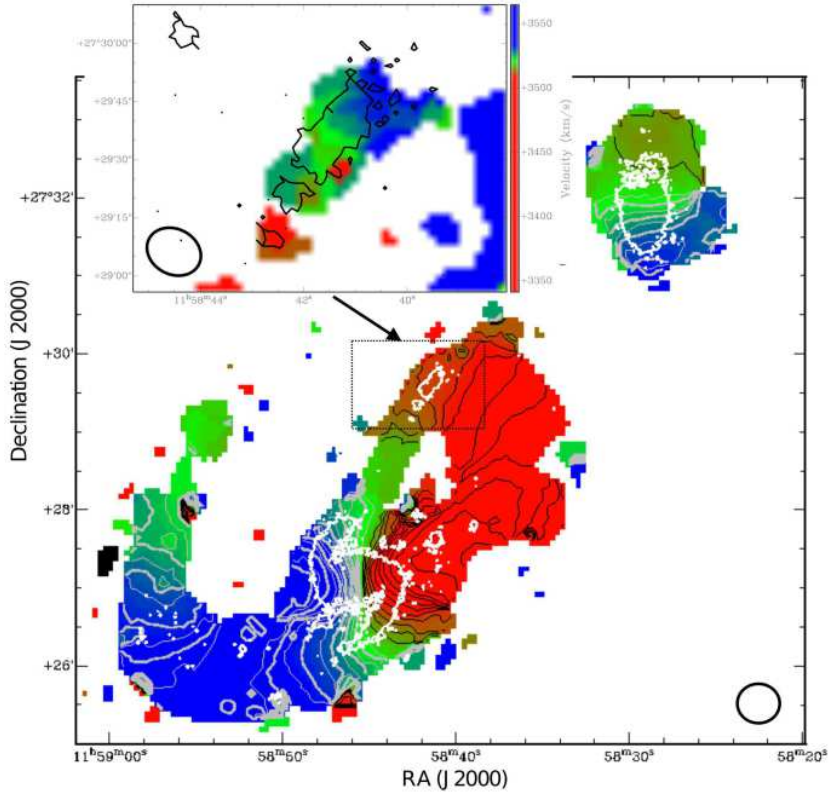}
\vspace{1cm}
\caption{\textbf{Arp\,305: \textcolor{black}{\textbf{\textit{Main figure}}} \hi\ velocity field } from the GMRT low resolution cube for \hi\ emission $>$ 3 $\sigma$.  The areas in red \textcolor{black}{with}  black contours   have velocities $>$ the NGC\,4017 systemic velocity of 3439 \km\  and areas in blue \textcolor{black}{with} grey  contours have velocities $<$ 3439 \km. The contours are in steps of 7 \km.  The thick white contour is  from the FUV (\textit{GALEX}) image. The  GMRT low resolution beam (31.8$^{\prime\prime}$ $\times$ 29.5$^{\prime\prime}$) is shown in bottom right corner. \textcolor{black}{\textbf {\textit{Inset }}Zoom in on the Bridge TDG region showing velocity field of the GMRT high  resolution \hi\ cube for emission $>$ 3 $\sigma$. \textcolor{black}{The GMRT high resolution beam (14.3.$^{\prime\prime}$ $\times$ 11.8$^{\prime\prime}$) is shown in bottom left corner.}}  }
\label{fig4}
\end{center}
\end{figure*}

In the \textcolor{black}{channel maps (Figures 8 and 9)  \hi\ emission is detected for NGC\,4017}   in the  velocity range  3292 \km\ to 3595 \km. Its  GMRT  V$_{HI}$ = 3439$\pm$ 7 \km\   \textcolor{black}{agrees well  with the } heliocentric optical radial velocity \textcolor{black}{for} NGC\,4017  (3449$\pm$2 \km).  \textcolor{black}{The GMRT} W$_{20}$  =  258$\pm$7\km\  is also similar to the  W$_{50}$ \textcolor{black}{ =} 253$\pm$5 \km\ from single dish on--line data \citep{haynes11}. The \textcolor{black}{NGC\,4017}  iso--velocity contours in the \hi\ velocity field (Figure \ref{fig4}) also show a fairly regular rotation \textcolor{black}{pattern}   (PA = 111\degree).  A position velocity (PV) diagram for a cut along the \textcolor{black}{NGC\,4017}  major  axis  (Figure \ref{fig6}), shows the \hi\ line centre to be 3447 $\pm$ 7 \km\ and maximum rotation velocity to be $\sim$ 150 $\pm$ 7 \km\ \textcolor{black}{(before inclination correction).} Two structures in the \textcolor{black}{NGC\,4017 PV diagram \textcolor{black}{(see Figure 6 -- right panel)} with \textcolor{black}{offsets} $>$ +0.6 arcmin and $<$ -1.0 arcmin, respectively,} are cuts through the SE and NW \hi\ tidal tails. The velocities in the \textcolor{black}{ \sett,} which in projection is an extension of the southern optical spiral arm, initially systematically decreases along  the tail to  a velocity $\sim$ 115 \km\ below  the NGC\,4017 systemic  velocity, \textcolor{black}{at which position}  the projected direction  begins  changing  northward. Beyond this position the \hi\ tail  velocities  systematically increase  reaching the NGC\,4017 systemic velocity at the end of the \hi\ tail. To the NW of NGC\,4017, from the \textcolor{black}{base of the \nwtt\  the \hi\ velocities increase systematically along the \nwtt\ all the way to  the \nwtb. Along the \nwtb\ itself  \hi\ velocities decline systematically in the direction of  NGC\,4017.  There is no clear kinematic break to  \textcolor{black}{distinguish} the \nwtt, \nwdr\ and the \nwtb.}  In fact  the kinematic continuity from \nwtt\ though  the \nwdr\ to \nwtb\ is puzzeling.  Modelling by \cite{hancock09} of the Arp\,305 system predicts the NW tidal bridge, but the origin of the \nwtt\  and \nwdr\ is unclear. \textcolor{black}{The} authors attribute them to ``material splashed out of the disks at closest approach". However, in terms of \hi\ mass, column density as well as kinematics, the NW \hi\ tidal tail appears to be an equally \textcolor{black}{unambiguous} and robust structure as the \nwtb.  %In section 4.1, we hypothesise that this could be remnant from a previous passage of NGC 4016 about NGC 4017.   

\textcolor{black}{Overall the \hi\ morphology and kinematics for NGC\,4017 suggest multiple   \hi\ structures in NGC\,4017: (i) a  regular rotating \hi\ disk  with a similar inclination to the optical galaxy (48\degree)  and  PA $\sim$111\degree; (ii) a \sett\ (iii)  \nwtt\  (iv) the \nwtb\ counterpart which is distinguishable  \textcolor{black}{ by \textcolor{black}{its} local column density peak\textcolor{black}{, although},  not kinematically;}  and the extensive \nwdr\ which is not  clearly distinguishable as a separate kinematic structure from the \hi\ bridge \textcolor{black}{or} NW tidal tail.}    %The velocities at the edges of the arms drops by $\sim$ 70 \km\ for the southern arm and by $\sim$ 50 \km\ for the NW arm. Additionally a shift in the PA of the \hi\ disk is also noticed. The PA of the un--warped central region of the \hi\ disk is estimated at 97$^{\deg}$, while the PA the \hi\ disk including the SE and NW arms is estimated to be 112 \degree.} 

\begin{figure*}
\begin{center}
\includegraphics[ angle=0,scale=.57] {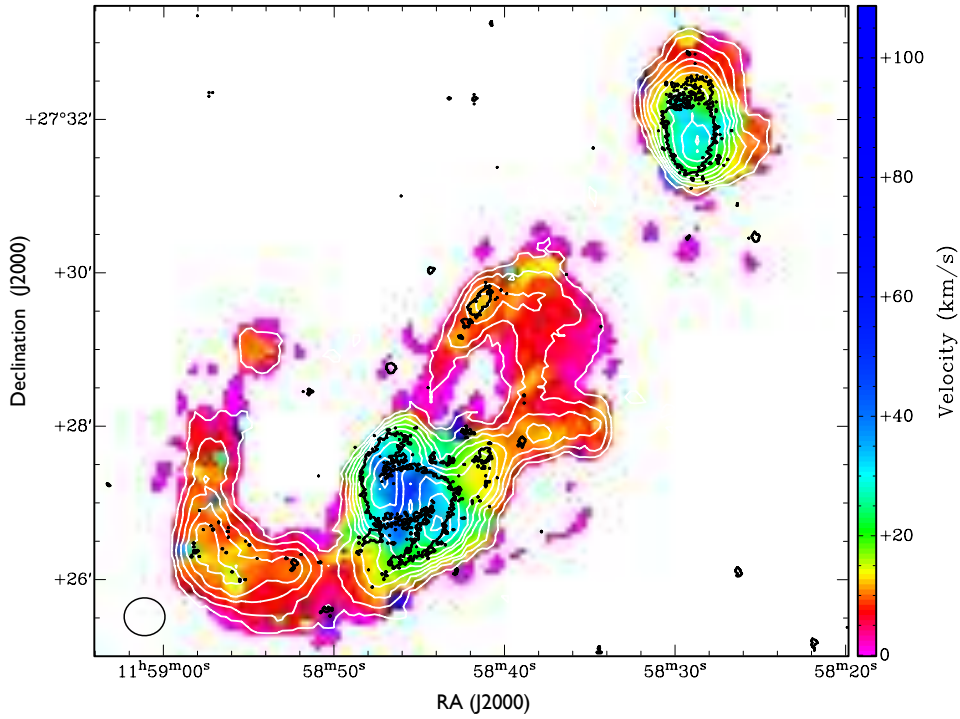}
\vspace{1cm}
\caption{\textbf{Arp\,305: GMRT velocity dispersion map from the low resolution cube.} \textcolor{black}{The white contours are from the low resolution \hi\ integrated map (Figure \ref{fig2}).} The ellipse at the bottom left indicates the GMRT synthesised beam \textcolor{black}{(31.8$^{\prime\prime}$ $\times$ 29.5$^{\prime\prime}$)}. }
\label{fig5}
\end{center}
\end{figure*}

\begin{figure*}
\begin{center}
\includegraphics[ angle=-0,scale=.45] {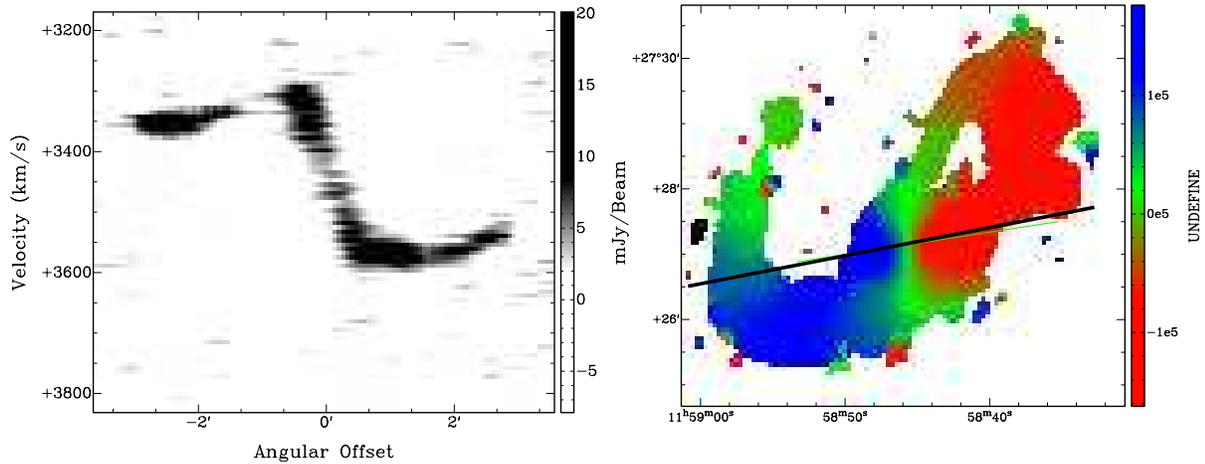}
\vspace{1cm}
\caption{\textbf{ NGC\,4017 PV diagram :} \textbf{\textit{Left }} PV diagram PA 291\degree. Positive  angular offset values are to the NW of the  kinematic  centre and negative values to the SE, with the angular \textcolor{black}{offset} scale in arcmin.  \textbf{\textit{Right}} GMRT NGC\,4017 \hi\ velocity field  showing the position of the PV \textcolor{black}{cut in} the left--hand panel. }
\label{fig6}
\end{center}
\end{figure*}

\begin{figure}
\begin{center}
\includegraphics[ angle=-0,scale=.47] {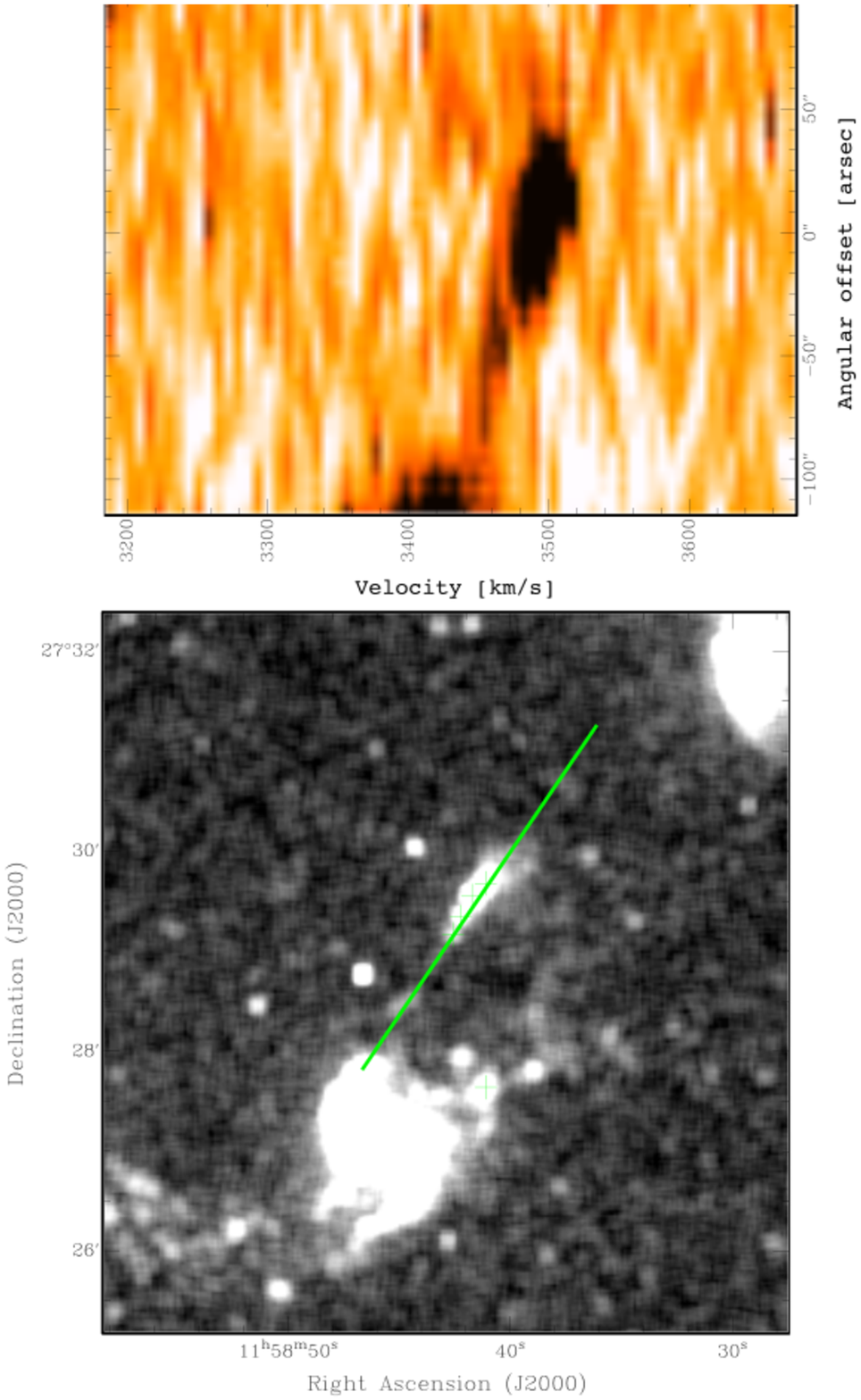}
\vspace{1cm}
\caption{\textbf{ Bridge TDG PV diagram :} \textbf{\textit{Top}}  PV diagram cut from the low resolution cube along the axis of the 4  FUV knots in the Bridge TDG. \textcolor{black}{ Negative angular offsets are in the SE direction}. \textbf{\textit{Bottom }} FUV image \textcolor{black}{\textit{(GALEX)}} showing the position of the PV cut (PA 146\degree) shown in the top  panel.   }
\label{fig7}
\end{center}
\end{figure}

\section{Discussion}
\label{dis}

\subsection{Interaction dynamics}
\label{dis_dyn}

%since the most recent NGC\,4017/NGC\,4016 interaction

In \hi,   Arp\,305 shows strong  tidal interaction signatures including,  the \sett, \nwtt\ and \nwtb.  Arp\,305 specific modeling  for a  prograde interaction by  \cite{hancock09} predicted  stellar counterparts to the \sett\  for  NGC\,4017 and the NW tidal bridge between the pair. Both tidal features are also predicted by  generalised modelling of a late-type galaxy undergoing a prograde  interaction with a minor companion \citep{oh08}. However, Arp\,305 also exhibits extended \hi\ debris (i.e., the \nwtt\  and the \nwdr) that  are  not predicted by either the \cite{hancock09} or \cite{oh08} models. The fact that these features are seen  in \hi\ and are bright in UV, but lack  optical counterparts, suggests these  features  are young  interaction debris.   \hi\ kinematic and morphology perturbations in the GMRT maps are more severe on the western side  of the NGC\,4017 disk, indicating NGC\,4016's closest approach  occurred there.  A possible explanation for the \nwtt\ is that it  was  \hi\ tidally drawn out of the NGC\,4017 disk during the pericentre approach of NGC\,4016.  \cite{hancock09} attributes this structure (\nwtt) to the material ``splashed out" during NGC\,4016's closest approach.  However,  in terms of \hi\ mass, column density as well as kinematics, the  \nwtt\  is as well defined  a structure as the \nwtb. Additionally the  \nwdr\ in the WSRT map, which is only partially recovered by the GMRT map, has an enormous  extent.  Also the continuous kinematic gradient  from  the \nwtt\  along the kinematic major axis of NGC\,4017 (Figure \ref{fig3}) until it merges with the \nwtb\  is inconsistent with the disturbed kinematics  expected in  ``splashed out" debris. In the absence of a better model we of course cannot make any robust claim. Alternative explanations could be  that (i) the \nwtt\ is a tidal feature attributable to the close approach of  NGC\,4016 on that side of the galaxy and (ii) at least part of the \nwdr\ is   \hi\ debris from an earlier encounter between the pair. Similar massive \hi\ structures have been previously reported in M\,51 type systems which could not be explained by modelling of a single encounter and were speculated to  originate from multiple passage encounters \citep{byrd,salo}.

\cite{oh08} used simulations to study the enhancement  of the spiral features and the duration of visibility  in disk galaxies following an interaction  with  a perturber. The similarity in the baryonic masses of NGC\,4017 ($\sim$ 4.0 $\times$ 10$^{10}$ \msolar) and the disk galaxy  used in  the Oh simulations ($\sim$ 5.2 $\times$ 10$^{10}$ \msolar) allows us to use the \cite{oh08} model to understand the physical properties of NGC 4017 (acknowledging  that the NGC\,4017 orbital parameters are poorly constrained in comparison with  the Oh simulations). \textcolor{black}{To acess the strength of the tidal interaction between  \textcolor{black}{simulated} galaxy pairs  and understand interactions with  $S$ $< 0.3$, i.e., moderately strong interations,  \cite{oh08} used a  tidal strength parameter}:\\ $S$ = ($\frac{M_p}{M_g}$) ($\frac{R_g}{R_{peri}}$)$^3$ ($\frac{\Delta T}{T}$) (their Eqn. 3).  \textcolor{black}{We can esitimate, for  NGC\,4017, the  perturber to  galaxy mass ratio ($\frac{M_p}{M_g}$) } \textcolor{black}{at}  $\sim$ 3, \textcolor{black}{a  typical value for}  interacting pairs with TDGs. \textcolor{black}{For NGC\,4017  we do not have any} \textcolor{black}{obervational} constraint for the  peri--centre to galaxy radius ($\frac{R_g}{R_{peri}}$) \textcolor{black}{ratio} or the perturber  angular speed relative to stars at the galaxy edge ($\frac{\Delta T}{T}$) . \textcolor{black}{Additionally, the \cite{oh08} simulations reveal} that the tidal tails  dissipate rapidly after reaching  their \textcolor{black}{ visiblity maxiumum,  1.4 to 2.5 $\times$ 10$^8$ yr following} the interaction. \textcolor{black}{For} S $>$ 0.3  \textcolor{black}{interactions},  \textcolor{black}{the  timescale for  tail dissipation may extend to $\sim$ 1 Gyr and  tail  fragmentation may lead  to TDG formation} \citep{barnes92,oh08}. \textcolor{black}{Here, using the Oh models as well as other observational evidence, we make an effort to constrain the time since the most recent NGC\,4017/NGC\,4016 interaction. Following are our four sources of evidence (i) for galaxies with  total baryonic masses of the  order of NGC\,4017, \hi\  morphological perturbation signatures from a full merger} only remain identifiable for a maximum of  $\sim$ 4  \textcolor{black}{$\times$} 10$^{8}$ yr to  7 $\times$ 10$^{8}$ yr   \citep{holwerda11}. We find the \hi\ morphology in NGC\,4017 to be strongly disturbed, i.e, well above the Holwerda \hi\ merger signature threshold. Assuming \hi\ perturbations from a full merger would be \textcolor{black}{of a similar magntude}  to the pre--merger interaction  observed in NGC\,4017,    \textcolor{black}{ it seems reasonable to conclude that the NGC\,4017} perturbation occurred well within the \cite{holwerda11} timescale upper limits; (ii)   NGC\,4016 and NGC\,4017 have a projected separation of  372 arcsec (90  kpc). \textcolor{black}{If we assume a} separation velocity of \textcolor{black}{ $\sim$ 212 \km, i.e. twice the \textcolor{black}{USGC} U435 group velocity dispersion,} \textcolor{black}{it implies   the  time  since their closest approach was} $\sim$ 4.1  $\times$ 10$^{8}$ years; \textcolor{black}{(iii) Using an inclination corrected rotational velocity of 201 \km\ , the time for a single rotation\footnote{ T$_{rot}$ [Gyr]= 6.1478 r/ V$_{rot}$ , where r = the optical radius [kpc] and V$_{rot}$= 0.5 $\Delta$V  [\km]/ sin(i).} of NGC 4017 is $\sim$ 0.6 $\times$ 10$^{9}$ yr. Following the \cite{oh08} simulations, we make the assumption that the bridge was formed along an axis joining NGC 4016 to NGC 4017 at the time of their closest approach and this point of closest approach has since rotated and reached its current location. From the orientation of the system and its \hi\ morphology and kinematics, it seems that the point of closest approach was the western edge of the NGC 4017 disk where the ``splashed out" material is visible.  Relative to the optical center in a  SDSS \textit{g} -- band de--projected   image\footnote{based on  PA = 111\degree\ and inclination = 48.2\degree} of NGC\,4017, the western disk edge  is offset by $\sim$ 262\degree\    in an anti--clockwise direction from the bridge. This implies that the NGC \,4017 disk has rotated  $\sim$ 262\degree\ since the bridge was formed and  we estimate the  time  since the bridge was formed at $\sim$ 4 $\times$ 10$^{8}$} yr \footnote{262\degree/360\degree\ $\times$ 0.6 $\times$ 10$^{9}$ yr = 4 $\times$  10$^{8}$ yr}; and  (iv) A comparison of the optical morphology of  NGC\,4017 and the  \cite{oh08} simulations (their Figure 1), shows a good agreement around  \textcolor{black}{t = 0.3 and 0.4} Gyr. \textcolor{black}{While the uncertainties for each of these timescales} are large, they sort of indicate that the \textcolor{black}{interaction took place} within the last \textcolor{black}{$\sim$ 4 } $\times$ 10$^8$ yr.   This timescale  \textcolor{black}{agrees well with the 3.8 $\times$ 10$^{8}$ yr} from the  \cite{hancock09} \textcolor{black}{modelling.} Even $\sim$ 4  $\times$ 10$^8$ yr after the \textcolor{black}{most recent}  pericentre approach   strong \hi\  \textcolor{black}{morphological and} kinematic perturbation signatures \textcolor{black} from the interaction remain clearly \textcolor{black}{observable}.  The  prominent optical tidal features of NGC\,4017 in this time frame \textcolor{black}{is consistent with}  the \cite{oh08} \textcolor{black}{simulations. This together with} the presence of TDG candidates  \textcolor{black}{support an agrument in favour of an }  \textcolor{black}{$S <$} 0.3 interaction. 

%and that the most optically and \hi\ disturbed part of the NGC\, 4017 disk edge was produced by that interaction, but contined rotation of the disk since then has displaced the disturbed region to its current location at the western disk edge.

\subsection{ Star formation and \hi\ column densities}
\label{res_sf}
A rich  \textcolor{black}{array of  star formation (SF) activity \textcolor{black}{in Arp\,305 is indicated by the emission detected in  \textcolor{black}{UV (\textit{GALEX})} images.} Extended areas of SF  are detected in FUV beyond  the optical   disks of  both NGC\,4016 and NGC\,4017 (Figure \ref{fig2}).  \cite{hancock09}  carried out a detailed study of SF in the tidal features of Arp\,305 using UV (\textit{GALEX}) data. \textcolor{black}{Those} authors identified  45 isolated SF clumps  within the extragalatic  tidal debris, including \textcolor{black}{in} four TDG candidates, see Tables 3 and 4 of \cite{hancock09}. \textcolor{black}{For NGC\,4017 we} explore the relationship between the Hancock extra--galactic FUV clumps and the \hi\ debris in which they are projected, i.e.,  in the extended \hi\ debris of the NW tidal tail, NW diffuse region and SE \hi\ tidal tail as  marked in \textcolor{black}{Figure \ref{fig3}}.  \textcolor{black}{Because of the GMRT's lower spatial resolution ($\sim$ 30$^{\prime\prime}$) compared to ultraviolet data from GALEX ($\sim$ 5$^{\prime\prime}$),} we study the \textcolor{black}{aggregated   star formation behaviour in each of} the \textcolor{black}{zones} marked   A to G in Figure \ref{fig3}. Table \ref{table4a} sets out the \hi\ column densities, \textcolor{black}{SFR and FUV clumps within} each SF region.  As noted in section \textcolor{black}{\ref{res_morph_mas}, extra--galactic tidal  \hi\ debris was detected at the projected positions of only two of the four TDG candidates}. SF zones A and D \textcolor{black}{correspond to the Bridge TDG  and TDG\,1 candidates respectively. Other SF zones projected within the \hi\ debris are centrally concentrated  star forming  clumps (zones B, C and G) and zones of multiple small faint clumps (zones E and F).}} 

\begin{table}
\centering
\begin{minipage}{150mm}
\caption{NGC\,4017 SF zone properties}
\label{table4a}
\begin{tabular}{llllll}
\hline
Zone&ID&SFR&\hi\ column&FUV\\
&&&density&clump\footnote{\textcolor{black}{From \citep{hancock09}.}}\\
&&[\msolaryr]&[\textcolor{black}{$\times$} 10$^{20}$ cm $^{-2}$]\\
\hline
A& Bridge TDG &0.200&4.1&12,13, \textcolor{black}{15,16} \\
B& &0.009&1.7& 14\\
C& &0.015&2.9& 42 \\
D& TDG 1&0.010&6.4 &11 \\
E& &0.020&5.3 & 37,38, \textcolor{black}{43}, 40\\
F& &0.011&4.1 &  \textcolor{black}{39}, 44 \\
G& &0.009&0.5 & 36 \\

\hline
\end{tabular}
\end{minipage}
\end{table}
%There are three SF zones identified in \citep{hancock09} which are not shown in Figure X. This is because they do not have \hi\ counterparts and in absence of any spectroscopic data it is dificult to claim them as part of the Arp 305 pair. 

 \cite{hancock09}  notes multiple strong star formation sites  over an extensive area of \textcolor{black}{the} SE tidal tail region and their model of the Arp\,305 interaction also predicts high gas densities in those strong star forming regions of the SE tidal tail. Indeed,  apart from the main galaxy disks of NGC\,4016/7, the SE tidal tail has  highest GMRT \hi\ column densities.  However, total  SFR in  SE tidal tail \textcolor{black}{(0.04 \msolaryr) is  an order of magnitude}  lower than in the \textcolor{black}{Bridge} TDG candidate \textcolor{black}{(0.2 \msolaryr)}.  The  SF zones  (A to G) are  projected  within  \hi\ with column densities ranging from  0.5 to 6.4  \textcolor{black}{$\times$} 10$^{20}$ cm $^{-2}$.  \textcolor{black}{However,} the \hi\ local maxima are not necessarily spatially correlated with the individual SF zones.  It is worth mentioning here that several studies have been conducted in the past to ascertain the critical \hi\ column density that triggers star formation in galaxies as suggested by \cite{kennicutt89}. Earlier studies, eg.  \cite{skillman}, found the limit to be $\sim$ 10$^{21}$ cm $^{-2}$ (for a spatial resolution of 500 kpc). A more recent study in the outskirts of the main galaxy disks and tidal debris found the limit to be $\sim$ 4 $\times$ 10$^{20}$ cm $^{-2}$  over a spatial resolution of about 1 kpc \citep{maybhate07}. The spatial resolution plays a crucial role as the quoted \hi\ column density values can change with the synthesised beam size. In \textcolor{black}{the} case of Arp\,305, the spatial resolutions we reach are $\sim$ \textcolor{black}{7} kpc with the low resolution map and $\sim$ \textcolor{black}{3} kpc with the high resolution map.  While this prevents us from drawing any firm conclusion about star \textcolor{black}{formation,  above} or below the critical column density regions in the Arp\,305 system, the wide range of \hi\ column density regions hosting star formation in Arp\,305  reaffirms that  a critical \hi\ column density may be a necessary criteria but not a sufficient one to initiate star formation \citep{begum}.

Areas of higher velocity dispersion in the extragalactic \hi\ debris in the GMRT \hi\ velocity dispersion map (Figure \ref{fig5})   correlate well with the SF zones (A-G).     For the SF zones projected \textcolor{black}{within}  the SE and NW \hi\ tidal tails and \nwtb, the \textcolor{black}{velocity} dispersion values range between 10 \km\ to 15 \km, higher than the usual 7 \textcolor{black}{\km\ to 8 \km\ in the non--star} forming areas of the \textcolor{black}{extragalactic \hi\ debris}.  \textcolor{black}{Within the optical  disks of the two galaxies, \hi\ } \textcolor{black}{velocity} dispersions are higher, $\sim$ 15\km\  to 40 \km.  \textcolor{black}{This is consistent with  \cite{mullan13} who found compact star clusters in \hi\ tidal tails are preferentially located in \hi\ regions with column densities  $>$ 4.6 $\times$ 10$^{20}$ cm $^{-2}$  (1 kpc resolution) and the highest  \hi\ velocity dispersions. \cite{mullan13}  also argues  higher \hi\ velocity dispersion is a condition for SF in the tidal tails, rather than consequence of the SF.} The star formation rates \textcolor{black}{(SFR)}, estimated from the FUV fluxes \citep{hancock09} \textcolor{black}{for} the \textcolor{black}{SF zones}   A to G are  0.2, 0.009, 0.015, 0.010, 0.020, 0.011 and 0.009 M$_{\odot}$ yr$^{-1}$, respectively. The \textcolor{black}{highest  SFR in zone  A \textcolor{black}{contains}  the bright star forming Bridge TDG candidate, consisting  of four strong SF clumps.}  \textcolor{black}{The Bridge} TDG candidate \textcolor{black}{is discussed further } in section \ref{dis_tdg}.

\subsection{\textcolor{black}{Arp\,305 Bridge TDG candidate} } 
\label{dis_tdg}
\begin{figure*}
\begin{center}
\includegraphics[ angle=0,scale=.65] {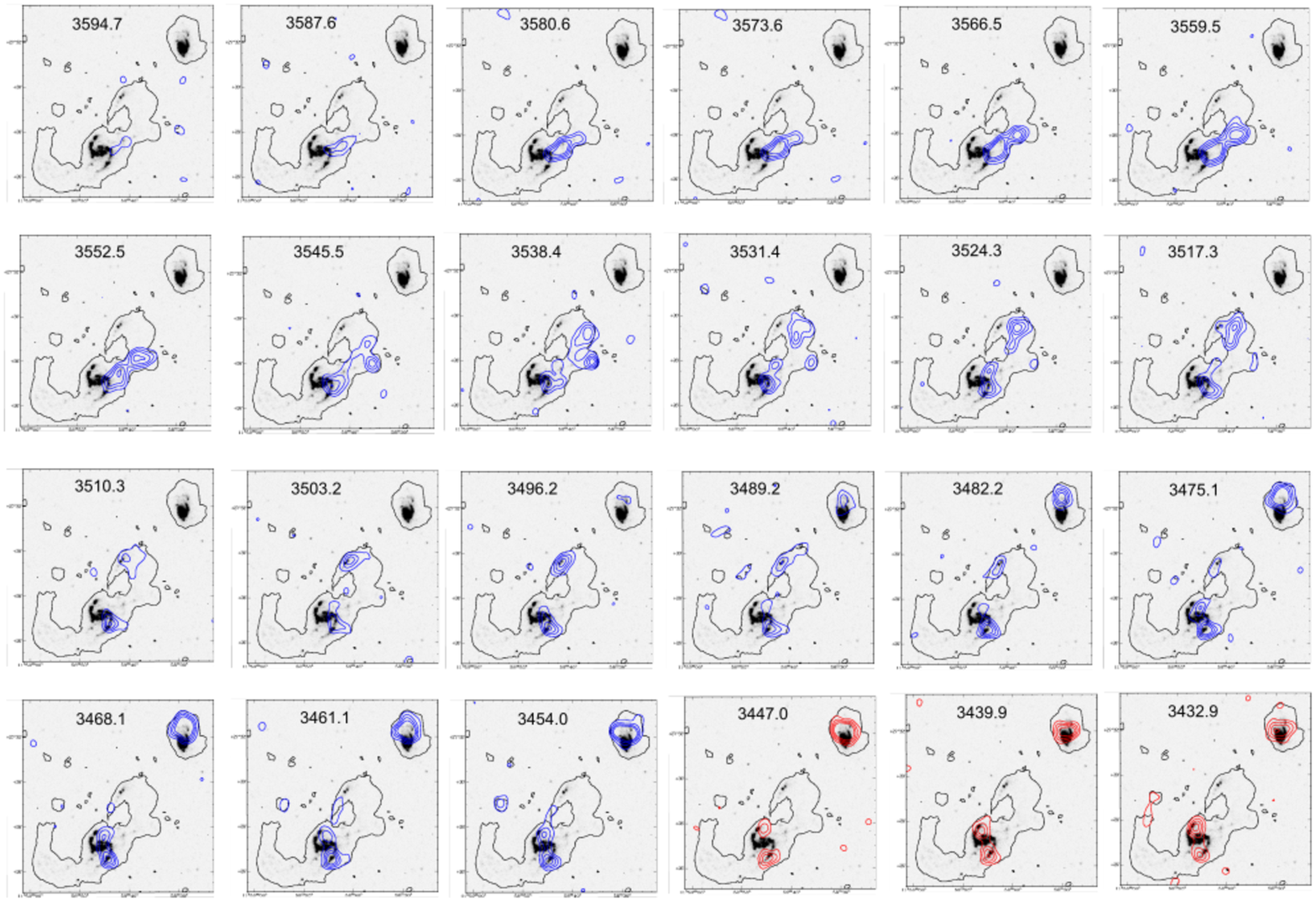}
\vspace{1cm}
\caption{\textbf{Arp\,305 \hi\ channel maps \textcolor{black}{from the low resolution cube -- part A.}} \textcolor{black}{Each channel's velocity in \km\ is shown at the top of each panel with  the channel separation $\sim$ 7 \km. The  contours indicate \hi\ emission, blue contours for \hi\ emission in channels with velocities higher than the pair's mean optical heliocentric velocity of 3445 \km.  The \hi\ contour colour changes from blue to red at 3445 \km\ for channels with velocities below this value. The contour levels are 1.2 mJy $\times$  (3, 5, 7, 9). The black contours are the lowest contour from the low resolution \hi\ velocity integrated map shown in Figure \ref{fig2}.  A continuation of the channel maps is presented in Figure \ref{fig9}.} }
\label{fig8}
\end{center}
\end{figure*}

\begin{figure*}
\begin{center}
\includegraphics[ angle=0,scale=.65] {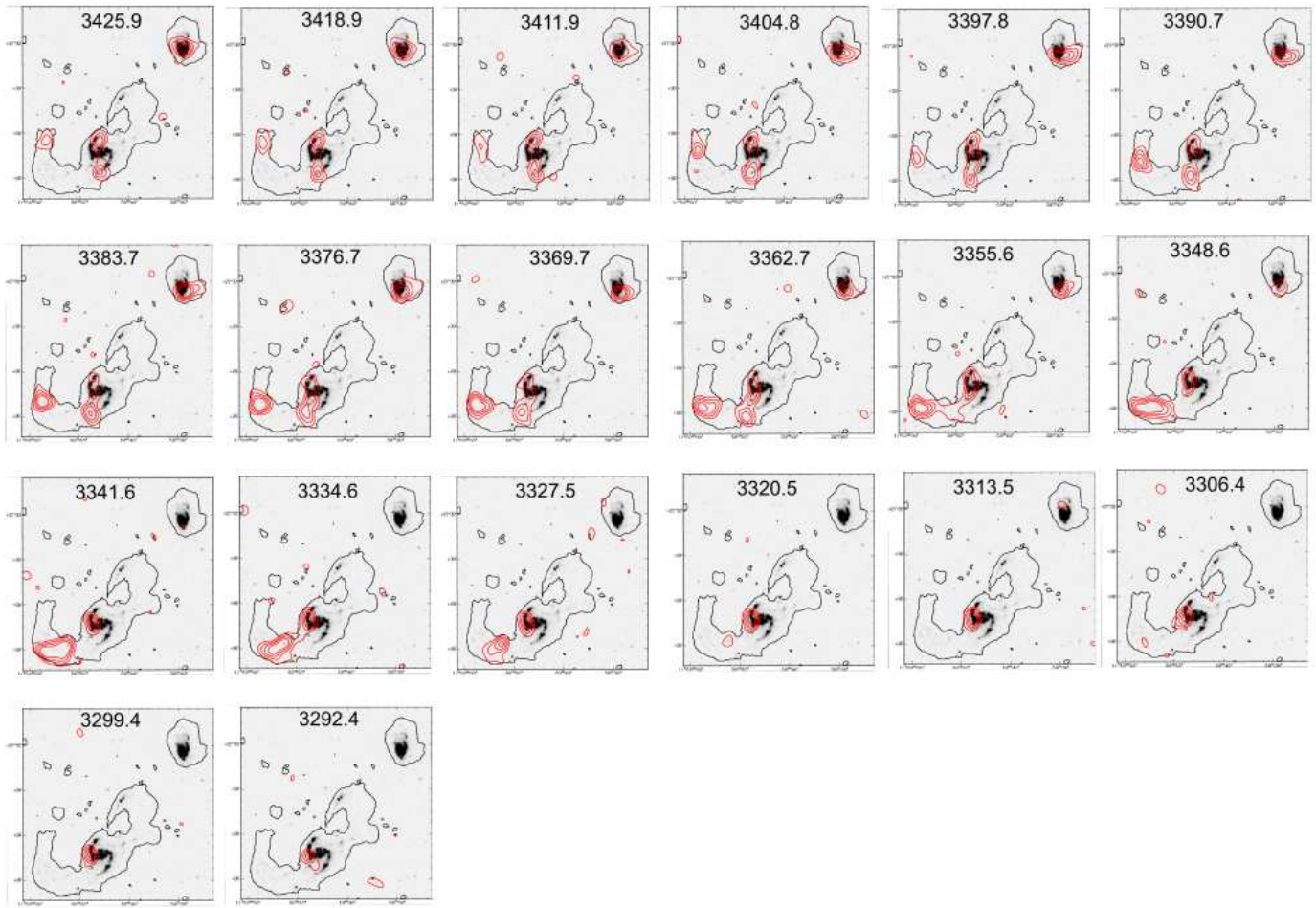}

\vspace{1cm}
\caption{\textbf{Arp\,305 \hi\ channel maps \textcolor{black}{from the low resolution cube -- part B.}.}  The description of the plot is given in the caption to  Figure \ref{fig8} caption. }
\label{fig9}
\end{center}
\end{figure*}

\textcolor{black}{Validation of a TDG candidate usually requires} a  combination of evidence linking  the candidate to the interaction between its parents, its metallicity, its stellar population  and \textcolor{black}{ gas disk rotation  signatures from \hi, CO or \halpha\ velocity fields}. The Bridge TDG candidate  emits strongly at  ultraviolet (UV) wavelengths and has  one of bluest   FUV -- g colours ($\sim$0.25) amongst  TDGs  in the \textcolor{black}{TDG sample studied by \cite{hess}, indicating strong recent SF. Within the \nwtb\ the  \hi\ column density maximum in the low resolution map  (4.1 $\times$ 10$^{20}$ atoms  cm$^{-2}$)  is projected at the position of the UV clumps 12 and 13 from \cite{hancock09}, Figure \textcolor{black}{ \ref{fig3}} -- right panel. \textcolor{black}{An \hi\ } spectrum for the Bridge TDG was extracted \textcolor{black}{from}  the low resolution cube, centred on its \hi\ column density maximum, which includes}  the projected positions of the  four Hancock UV clumps (numbered 12, 13, 15 and 16). This spectrum  provides an upper limit  for the \hi\ mass of the Bridge TDG ($\sim$ 6.6 \textcolor{black}{$\times$} 10$^{8}$ M$_\odot$).   The velocity field and a PV diagram cut from the low resolution cube, taken along ``tidal bridge " \textcolor{black}{major} axis \textcolor{black}{(Figure \ref{fig7})} shows \hi\ detected \textcolor{black}{  in the range of 3475 -- 3520 \km\  with a modest gradient and velocities increasing in the NW direction.} \textcolor{black}{The Bridge TDG's V$_{HI}$ = 3500$\pm$7\km\ is in good agreement with the velocities of the parent galaxies.} The velocity gradient is \textcolor{black}{also clearly seen} in the high resolution \textcolor{black}{velocity field \textcolor{black}{(inset in Figure \ref{fig4})}.} \textcolor{black}{The} PV diagram,  the channel maps  (Figures \ref{fig8} and \ref{fig9}) and the  \hi\ map \textcolor{black}{(Figures \ref{fig2} and \ref{fig3})} \textcolor{black}{reflect the concentration of \hi\ along the major axis of the tidal bridge in the vicinity of UV clumps 12, 13, 15 and 16, with the } highest \hi\ column density  at the projected position of UV clumps 12 and 13. There is also a local maxima for  velocity dispersion of $\sim$ 14 \km\ at this position (Figure \ref{fig5}).

In optical/ UV \textcolor{black}{images the Bridge TDG has an} $\sim$ 11 kpc length and the low and high resolution GMRT synthesised beams  ($\sim$32$^{\prime\prime}$ and  $\sim$14$^{\prime\prime}$) samples it at $\sim$ 7 kpc and 3 kpc respectively. The  spectrum shows an \hi\   line width of \textcolor{black}{$\sim$} 30 \km\ with  a systematic gradient over $\sim$ 50 \km\ across  its major axis, the velocity resolution being $\sim$ 7 \km. These values agree well with those TDGs reported in the literature, for which  a velocity gradient has been determined. In a recent work, \cite{lelli}, found signs of regular velocity gradient in  six bona--fide TDGs. Using a velocity resolution of 7 -- 10 \km\, and a spatial resolution of about 2 to 3 beams across the major axis, they report gradients between 25 \km\ \textcolor{black}{and} 80 \km. \textcolor{black}{A similar \hi\ velocity gradient of $\sim$ 30 \km\ to 40 \km\ was reported for a TDG candidate  in the Leo triplet \citep{nikiel}.}    While the \hi\ line width and velocity gradient estimates  \textcolor{black}{for the Bridge} TDG are consistent with those found in the literature\textcolor{black}{, it remains  unclear whether the} velocity gradient represents the intrinsic rotation of the Bridge TDG, or just  the gradient within the \hi\ debris.

Using the \hi\ spectrum from the low resolution cube for the \textcolor{black}{Bridge} TDG, we estimated its dynamical mass \textcolor{black}{(M$_{dyn}$)} to be 7 \textcolor{black}{$\times$ 10}$^{8}$\textcolor{black}{ \msolar.}   Based on their best fit models and scaling to the SDSS r--band flux, the stellar mass of the \textcolor{black}{Bridge} TDG was estimated at \textcolor{black}{1--7 $\times$}  10$^{6}$ M$_\odot$ by \cite{hancock09}, \textcolor{black}{giving}  \textcolor{black}{a $\frac{M_{HI} + M_*}{M_{dyn}}$ \textcolor{black}{ratio of} $\sim$ 1. This \textcolor{black}{ratio} is consistent with the absence of a substantial dark matter component and \textcolor{black}{is typical of the ratio} found for  \textcolor{black}{validated}  TDGs.  However, the following factors together make the estimate highly uncertain\textcolor{black}{: \textcolor{black}{(i)} the} TDG is embedded in an \hi\ debris and thus its spectrum  can be contaminated by foreground and background emission \textcolor{black}{(ii)} it is impossible to distinguish the extent of the \hi\ disk of the TDG from the general bridge emission  \textcolor{black}{(iii) }assuming the Bridge TDG was formed during the last encounter between the pair ($\sim 4 \times $ 10$^8$yr), it seems probable that there has been insufficient time for the Bridge TDG to virialise \citep{flores16}. All these factors make the calculated M$_{dyn}$ highly uncertain and therefore while we present our estimates here, we choose not to make any strong claims on the dark matter content of the TDG based on it.  } 

%It is  observationally} challange to establish a candidate TDG as a genuine one from its \textcolor{blue}{\hi\ } kinematics.
%Therefore while we discuss the \hi\ and star formation properties of the \textcolor{blue}{Bridge} TDG,  we also briefly discuss  why based solely on the \hi\ observations we cannot claim the \textcolor{blue}{Bridge} TDG candidate to be a real TDG. 
The \cite{hancock09} model of the stellar component of the pair interaction shows the development of \textcolor{black}{a} tidal bridge between NGC\,4017 and NGC\,4016,  with SF activity near the centre of the bridge and the bridge base near NGC\,4017. The authors suggest, ``material balanced between the two galaxies", NGC\,4017 and NGC\,4016, \textcolor{black}{collapsed under its own gravity and gave} rise to the TDG, i.e., the stellar debris \textcolor{black}{provided} the seed for accumulation and of gas debris which in turn \textcolor{black}{fuelled} SF in the TDG. Potentially  a kinematic rotation signature could  confirm a TDG  candidate as an independent galaxy, rather than  just an accumulation of  SF zones. However, in  this case this is not feasible as the Bridge TDG is embedded  in the bridge  \hi\ debris and  the GMRT spatial resolution is too poor to distinguish the  TDG candidate's  intrinsic kinematics from the kinematics of the \hi\ debris.   Due to their  relatively higher metallicity than standard dwarf galaxies, the probability of detecting CO emission lines is higher in TDGs \citep{braine} which could overcome the spatial resolution issue. Moreover, since a TDG's molecular gas is predicted to be formed in--situ \citep{braine}, the  molecular gas disk is expected to be more localised to the TDG than \hi.

 \begin{table*}
\centering
\begin{minipage}{190mm}
\caption{\textcolor{black}{ \hi\ and SF properties of the TDG candidates }}
\label{table_5}
\begin{tabular}{llrrrrrr}
\hline
%Frequency & Observation  & Phase      & Phase cal    &  $\tau$    & Bandwidth &rms (per channel  & beam size   \\ 
System& TDG             & \hi\ column    & \hi\ map       & \textcolor{black}{Estimated time}      &Pair M$_*$          & SFR             & SFE \\ 
      &candidate        & density        & resolution     &\textcolor{black}{ since interaction}    & ratio      &       & \\
      &                 & [10$^{20}$ atoms cm $^{-2}$] & [arcsec] &[Gyr] & & [\msolaryr] & [\msolaryr/\msolar] \\
\hline

Arp\,65  &\hi\ maxima    & 8.3    & 23 & 0.2   & 1:3.0  & --  &  -- \\
Arp\,181&  TDG           & 9.9     &10 & -- & 1:3.6 &-- \\
Arp\,202&  TDG           & 7.5     &23 & 0.4   &  1:1.4 & 0.04 & 3.9$\times$ 10$^{-10}$ \\
%Arp\,305&TDG\,1&14.2$\pm$1.9&&30 &0.4&1:3.3\\
Arp\,305 & Bridge TDG    &4.1     &32 &0.4      &1:3.3  &0.20 & 3.0 $\times$ 10$^{-10}$ \\

\hline
\\

%Velocity\footnote{For Arp\,305 TDG1  mean velocity dispersion with a 30" region centred on UV clump 11.  For Arp\,305 Bridge TDG the mean velocity dispersion is from and an area bounded by FUV clumps 12 and 16}
%dispersion&

%$\pm$2.4&\textcolor{blue}{4.1}
\end{tabular}
\end{minipage}
\end{table*}

\textcolor{black}{An old stellar component in the Arp\,305 tidal bridge is predicted by the \cite{hancock09} modelling of the pair interaction. Like the tidal bridges in  \cite{smith07}, the Bridge TDG is detected in both the \textit{Spitzer} 3.6 $\mu$m  and 4.5 $\mu$m band images. In general, it is understood that both bands trace emission from stellar populations with ages $>$ 1 Gyr.  However,  it is known that both of these bands can be contaminated by emission from strong SF regions,  by up to 50\% from intermediate age stars (RSG and AGB)   and 22\% from dust   \citep{meidt12}. Additionally the  4.5 $\mu$m band suffers from  CO absorption \citep{meidt12}. We  estimated the  \textit{Spitzer} 3.6 $\mu$m  and 4.5 $\mu$m  flux densities  at 94.71 $\mu$J  and  95.26 $\mu$J,  respectively. The \textit{Spitzer }magnitudes (AB) for these bands are 15.7$\pm$0.3 and 15.9$\pm$0.3, respectively and the [3.6]--[4.5] colour is $\sim$ -0.2 mag. Unfortunately there are inconsistent  interpretations for this color in the literature. For example,  Figure 7 of \cite{smith05},  indicates emission from M0 III stars (red giants) have a  colour ($\sim$ - 0.15), close to    - 0.2, while \cite{querejeta15} states ``The expected color for an old stellar population of ages t $\sim$ 2 Gyr --12 Gyr is  - 0.2 $<$ [3.6]--[4.5] $<$ 0 ". Both Smith and Querejeta agree that this colour is not associated with strong dust emission. Also the  [3.6]--[4.5] colour of old ellipticals is negative \citep{peletie12}.  We conclude that the Bridge TDG is largely free from dust emission  and it is highly probable that its 3.6 $\mu$m  and 4.5 $\mu$m emission is principally  tracing  a stellar population formed before the latest interaction by the pair. If this is the case,   it is consistent with the stellar  debris seeding the TDG scenario. Optical spectroscopy could provide confirmation of this.  The stellar mass based on the Spitzer  3.6 $\mu$m  and 4.5 $\mu$m  flux is $\sim$ 4.1 $\times$ 10$^7$\msolar\ following \cite{eskew12}. This is higher than the stellar mass estimate by \cite{hancock09} of 1--7 $\times$ 10$^6$\msolar, but still an order of magnitude lower than the \hi\ mass of the TDG, making it a gas dominated system.}

Accepting the limitations of claims that can be made using the \textcolor {black}{currently available}  data, we find that the \textcolor{black}{ Arp\,305 Bridge} TDG   differs from our previous TDG \textcolor {black}{\hi\ detections in Arp\,202 and Arp\,181 \citep{seng14,seng13} in having  \textit{Spitzer} 3.6 $\mu$m  and 4.5 $\mu$m  counterparts, indicative of an old stellar \textcolor{black}{component}.  Additionally, Arp\,305 TDG \textcolor{black}{has strong }  UV emission \textcolor{black}{indicating} recent SF. This is consistent with a scenario where  central region of the tidal bridge, containing old stars  \textcolor{black}{originating in principal pair, \textcolor{black}{provided} the seed potential for the TDG to grow from infalling gas debris. This scenario is quite different from TDG candidates detected in Arp\,181 and Arp\,202. In those cases weak optical emission and \textcolor{black}{the} absence of \textit{Spitzer} NIR emission \citep{smith07} suggests an  insignificant old stellar component in \textcolor{black}{those}  TDGs and supports  a  scenario  where the TDG primarily forms from  gas debris collapse.} In Table \ref{table_5} we compare the \hi\ and star formation properties of  the TDG candidates and potential TDG  host debris we have studied so far.  \textcolor{black}{Arp\,181's} TDG has no published SFR from any band and Arp 65's high column density  \hi\ debris \textcolor{black}{ does not host a TDG} or detected  star formation activity \citep{seng15}.  Table \ref{table_5} shows \textcolor{black}{that while the} SFR of Arp\,305 is about an order of magnitude higher than Arp\,202, \textcolor{black}{the} star formation efficiency (SFE), defined as SFR per unit \hi\ mass, \textcolor{black}{is}  similar and is consistent with the low SFE trends of TDGs \citep{braine}. While \textcolor{black}{the comparison}   of TDGs in Table \ref{table_5} \textcolor{black}{is} inconclusive due to small sample size\textcolor{black}{. It} remains an open question \textcolor{black}{whether the presence of a substantial old  stellar population in a TDG significantly affects its SF history and in particular the timescale for TDG formation.}}

%And estimate the dark matter content and influence on TDGs makes it 

\section{Summary and concluding remarks}
\textcolor{black}{We have mapped the \hi\ in Arp\,305 interacting pair  with the GMRT. Our analysis of the \hi\ morphology and kinematics of the pair supports the conclusion in \cite{hancock09}  that the  most recent encounter between the pair occurred $\sim$ 4 $\times$ 10$^8$ yr ago. However, there  are \hi\ morphological and kinematic features NW of NGC\,4017  \textcolor{black}{, not found in models of  first encounters for interacting galaxy pairs,}  which may be remnants of  an earlier encounter between the two galaxies. \textcolor{black}{Similar features in M\,51 type systems are proposed as debris from earlier encounters.  }  The Arp\,305 system shows extended star formation in \textcolor{black}{its  tidal tails and bridge}. The \textcolor{black}{GMRT \hi\ maps  lack the spatial resolution for detailed studies of the correlation between individual SF zones and \hi\ column densities, although the extragalactic  SF zones in the Arp\,305 system are projected  at locations with a range of  \hi\ column densities with no specific bias towards higher column densities.}}

 \textcolor{black}{The \hi\ morphology and kinematic properties of the Bridge TDG candidate \textcolor{black}{include:} \textcolor{black}{\mhi\ } $\sim$  6.6 $\times$ 10$^8$\msolar\ and V$_{HI}$ = 3500$\pm$7\km\ \textcolor{black}{(in good agreement with the velocities of the parent galaxies)}. Additionally \textcolor{black}{the} linewidth of 30\km, \textcolor{black}{the} modest velocity gradient, \textcolor{black}{and} SFR of 0.20 \msolaryr\  add to the evidence favouring  the Bridge TDG candidate being  a genuine TDG. %\textcolor{red}{However, the spatial resolution of the GMRT \hi\ and doubts about the virialisation state of the \hi\  in the TDG makes the M$_{dyn}$ used in the calculation above and the implied dark matter  mass highly uncertain.  }  
A \textit{Spitzer} 3.6 $\mu$m  and 4.5 $\mu$m  counterpart with a [3.6]--[4.5] colour $\sim$ -0.2 mag  suggests a formation scenario containing a substantial old stellar population.  Future spectroscopic observations \textcolor{black}{for} this TDG are planned to confirm this formation scenario and provide the metallicity of the TDG. \textcolor{black}{Originating from }processed material, TDGs are expected to show higher metallicity compared to normal dwarf galaxies, making it  a key criteria  for the validation for TDG candidates.}

\label{summary}

\section{Acknowledgements}
TS acknowledges support for this project from the Funda\c{c}\~{a}o para a Ci\^{e}ncia e a Tecnologia (FCT)  grant No.SFRH/BPD/103385/2014. We thank the staff of the {\it GMRT} who have made these observations possible. The {\it GMRT} is operated by 
the National Centre for Radio Astrophysics of the Tata Institute of Fundamental Research. This research 
has made use of the NASA/IPAC Extragalactic Database (NED) which is operated by the Jet Propulsion Laboratory, 
California Institute of Technology, under contract with the National Aeronautics and Space Administration. This research has made use of the Sloan Digital Sky Survey (SDSS). Funding for the SDSS and SDSS-II has been provided by the Alfred P. Sloan Foundation, the Participating Institutions, the National Science Foundation, the U.S. Department of Energy, the National Aeronautics and Space Administration, the Japanese Monbukagakusho, the Max Planck Society, and the Higher Education Funding Council for England. The SDSS Web Site is http://www.sdss.org/.\textcolor{black}{This research made use of APLpy, an open-source plotting package for Python hosted at http://aplpy.github.com}.

\bibliographystyle{mn2e}
\bibliography{cig}

%%%%%%%%%%%%%%%%%%%%%%%%%%%%%%%%%%%%%%%%

%%%%%%%%%%%%%%%%%%%%%%%%%%%%%%%%%%%%%%%%

% The first contour is at 3 $\sigma$ above the noise in the channel,  equivalent to  a column density of x atoms cm$^{-2}$ with subsequent contours art x atoms cm$^{-2}$. 

\end{document}